\documentclass[10pt,english]{elsarticle}
\usepackage{mathptmx}
\usepackage[T1]{fontenc}
\usepackage[latin9]{inputenc}
\usepackage{geometry}
\usepackage{amsmath}
\usepackage{amsthm}
\usepackage{graphicx}
\usepackage{setspace}

\makeatletter
\theoremstyle{plain}
\newtheorem{thm}{\protect\theoremname}
\ifx\proof\undefined
\newenvironment{proof}[1][\protect\proofname]{\par
\normalfont\topsep6\p@\@plus6\p@\relax
\trivlist
\itemindent\parindent
\item[\hskip\labelsep\scshape #1]\ignorespaces
}{%
\endtrivlist\@endpefalse
}
\providecommand{\proofname}{Proof}
\fi

\@ifundefined{date}{}{\date{}}
\journal{xxxx}

\makeatother

\usepackage{babel}
\providecommand{\theoremname}{Theorem}

\begin{document}

\begin{frontmatter}{}

\title{\noindent Placebo Response is Driven by UCS Revaluation: Evidence,
Neurophysiological Consequences and a Quantitative Model}

\author[rvt]{Luca Puviani}

\ead{luca.puviani@unimore.it}

\author[focal]{Sidita Rama}

\address[rvt]{{\small{}Department of Engineering ``Enzo Ferrari'' , University
of Modena and Reggio Emilia, Via Vivarelli 10, int.1-41125 Modena
, Italy}}

\address[focal]{{\small{}MD at Local Health Unit of Modena, Via S.Giovanni del Cantone
23, 41121 Modena , Italy}}
\begin{abstract}
\noindent Despite growing scientific interest in the placebo effect
and increasing understanding of neurobiological mechanisms, theoretical
modeling of the placebo response remains poorly developed. The most
extensively accepted theories are expectation and conditioning, involving
both conscious and unconscious information processing. However, it
is not completely understood how these mechanisms can shape the placebo
response. We focus here on neural processes which can account for
key properties of the response to substance intake. It is shown that
placebo response can be conceptualized as a reaction of a distributed
neural system within the central nervous system. Such a reaction represents
an integrated component of the response to open substance administration
(or to substance intake) and is updated through ``unconditioned stimulus
(UCS) revaluation learning''. The analysis leads to a theorem, which
proves the existence of two distinct quantities coded within the brain,
these are the \emph{expected} or \emph{prediction outcome} and the
\emph{reactive response}. We show that the reactive response is updated
automatically by implicit revaluation lerning, while the expected
outcome can also be modulated through conscious information processing.
Conceptualizing the response to substance intake in terms of UCS revaluation
learning leads to the theoretical formulation of a potential neuropharmacological
treatment for increasing unlimitedly the effectiveness of a given
drug. 
\end{abstract}

\end{frontmatter}{}

\clearpage{}

\noindent \begin{flushleft}
\textbf{Introduction}
\par\end{flushleft}

\noindent What is the neural substrate of placebo response? Are there
neurons endowed with a special location or specific neurobiological
pathways that are necessary or sufficient for placebo response? Recent
approaches have attempted to narrow the focus, and a major insight
from the recent publications is that there seems not be a single neurobiological
or psychobiological mechanism which is able to explain the placebo
effect in general. Instead, different mechanisms exist by which placebo
(and nocebo) responses originate and exert their actions. Here, we
pursue a different approach. Instead of arguing whether a particular
neurobiological pathway, mechanisms, or group of neurons contributes
to placebo response or not, our strategy is to characterize the kinds
of neural processes (at the system-level) that might account for key
properties of the more general response to substance intake. With
this aim we start analyzing the response to (\emph{open} or \emph{overt})
pharmacological administration. Drug administration is said to be
\emph{open }(or \emph{overt},\emph{ }opposite to \emph{hidden} or
\emph{covert}) whenever the administration itself is correctly perceived
by a given subject; so that the individual is conscious of the administration
act, and, generally speaking, he/she might also be aware of the effects
that such a substance will determine on the organism (e.g. see \citep{Benedetti2008,Colloca2004}).
We emphasize three properties: pharmacological induced response is
\emph{integrated} (the response is given by the integral contribution
of different components, which are the \emph{active} pharmacological
effect, the \emph{reactive} contribution, due to \emph{self-induced}
response, which, in turn is governed by UCS revaluation, conditioning
and higher cognitive processing, such as beliefs), at the same time
it is highly \emph{differentiated} (one can experience any of a huge
number of different pharmacologically-induced response). Furthermore,
a third property is represented by the fact that the reactive response
mimics the active pharmacological effect (we denote this property
as \emph{reactive mimicking}). We first consider neurobiological data
indicating that neural processes associated with drug administration
response are integrated and differentiated, and, finally, that a distributed
system (denoted \emph{reactive system}) mimics the active pharmacological
response. We then provide mathematical tools describing the above
mentioned properties of the pharmacological response; this leads us
to formulate theoretical and operational criteria for determining
whether the activity of a group of neurons contributes to placebo
response and with which intensity. Moreover, the derived model suggests
a pharmacological strategy for increase unlimitedly the response to
a drug treatment (and hence the drug effectiveness), exploiting the
reactive system dynamics which governs the implicit placebo response.

\part*{Methods}

\section*{\noindent General Properties of (Overt) Drug Administration-Induced
Response}

\noindent Generally speaking, the effect of a drug administration
is due to the integration of different processes and components which,
together, exert their effect on a given organism. 

\begin{singlespace}
\emph{Integration - }Integration is a property shared by every response
induced by substance intake, irrespective of its specific pharmacological
target: Each response comprises a single overall effect that cannot
be decomposed into independent components. More specifically, drug
administration, and generally speaking substance consumption, exerts
an overall effect wich is perceived as unique, conveying pharmacological
effects, rewarding, expectations, beliefs and prior experiences (e.g.,
see \citep{Amanzio1999,Benedetti2008,Bingel2011}). A striking demonstration
is given by the decreased effectiveness of hidden treatments. In particular,
hidden drug administrations eliminate the psychosocial (placebo) component,
so that only the pharmacodynamic effect of the treatment (free of
any psychological contamination) can be accomplished. More specifically,
in the \emph{open versus hidden treatments }paradigm\emph{ }\citep{Amanzio1999,Benedetti2008,Benedetti2011,Colloca2004}
drugs are administered through hidden infusions by machines; such
infusions can be administered using computer-controlled infusion pumps
that are preprogrammed to deliver drugs at a desidred time. The crucial
factor is that the patients do not know that the drug is being injected.
The difference between the responses following the administration
of the overt and covert therapy is the placebo (psychological) component,
even though no placebo has been given \citep{Benedetti2008}. Furthermore,
in \citep{Amanzio1999} are reported experimental results which show
how the overall response to an analgesic pharmacological treatment
is due to the contribution of 1) prior conditioning, 2) expectations
and 3) active pharmacological effect; moreover, the three above mentioned
contributions are indirectly evaluated through the involvement of
different treatment groups in double-blind design, exploiting the
open versus hidden paradigm. 

\emph{Differentiation - }While the drug administration response is
given by an integral contribution of different components, there are
experimental evidences which show how the mammalian brain can finely
discriminate between administration (and/or the intake) of different
substances. For instance, neurons in rats can discriminate between
cocaine and liquid rewards \citep{Bowman1996,Carelli1994}, or between
cocaine and heroin \citep{Chang1998}, possibly even better than between
natural rewards \citep{Carelli2000,Schultz2000}. 

At this point it is useful to introduce some definitions. Provided
that a given drug administration represents an \emph{unconditioned
stimulus} (UCS), an \emph{active response} is defined as the response
determined by the active pharmacological effect of the UCS within
the central nervous system (CNS). Moreover, a \emph{reactive response
}is defined as any response which is determined by a self-induced
reaction within the CNS, without any active pharmacological contribution.
For instance, a conditioned response (CR) induced after some pairings
of an active pharmacological UCS and a neutral conditioned stimulus
(CS) corresponds to a reactive response, since the elicited response
is not substained by any pharmacological active effect. For this reason,
a reactive response could represent a specific form of placebo. Furthermore,
the UCS \emph{implicit learning} is defined as the implicit (or automatic)
UCS response (UCR) evaluation and re-valuation (i.e., UCR \emph{inflation}
and \emph{devaluation} or \emph{deflation}; e.g., see \citep{Hosoba2001,Puviani2016,Rescorla1974}).

\emph{Difference between (pharmacological) conditioning and implicit
UCS revaluation. }When a neutral CS is paired with an active drug,
the associative learning process is called classical conditioning
\citep{Pavlov1927}, (for instance when a tone is paired with the
self administration of cocaine in rats; \citep{Ito2000}); furthermore,
when a drug response (UCR) is evaluated and revaluated over trials
the learning process is represented by UCS revaluation \citep{Puviani2016}.
In other words, from one hand, classical conditioning learning occurs
whenever the CS and the UCS corresponds to different discriminable
elements, such that the organism can learn the simultaneously (or
causal) co-occurrence of both elements (in other words the ``association''
of two distinct elements is learned) \citep{Pavlov1927}. Instead,
on the other hand, in successive trials involving drug administration
there is a unique element (i.e. the drug administration itself, which
represents the UCS), which determines an UCR which, in turn, could
vary during successive administration trials; more specifically, if
the dose of a drug (i.e. the UCS active stimulation intensity) changes
over successive open administration trials, the organism updates the
evaluation of the experienced outcome through UCS revaluation (for
instance, in \citep{Hosoba2001,Rescorla1974,Schultz2013} experiments
related to UCR revaluation, independently from classical conditioning,
are reported; furthermore, a quantitative analysis of UCS revaluation
and its relation with classical conditioning are derived in \citep{Puviani2016}).
It is worth noting that, generally speaking, both classical conditioning
and UCS revaluation could occur during drug administration, since
the context and other cues could be associated as CSs to the drug
administration (UCS).

Finally, a generic induced response within the CNS can be represented
by the superposition of the activity of different neural populations;
more specifically, assuming that the CNS consists of $N$ different
neuronal populations, the response, denoted with the vector $\mathbf{y}$,
can be expressed as:

\begin{equation}
\mathbf{y}=\sum_{i=1}^{N}y_{i}\mathbf{v}_{i},\label{eq:base_model}
\end{equation}

where $y_{i}$ represents the i-th neuronal population activity\footnote{In particular, $y_{i}$ is a real quantity representing the product
between the mean number of elicited neurons and their mean firing
rates for the $i$-th neuronal population (with $i=1,2,...,N$); consequently,
$y_{i}$ takes on a positive (negative) value if the response produces
an increase of (a decrease or inhibition of) the activity for the
$i$-th population, and is equal to zero whenever the response does
not involve any adjustment for the baseline activity of the population.} and $\left\{ \mathbf{v}_{i};i=1,2,...,N\right\} $ represents a set
of versors, being associated with different neuronal populations,
which form a complete basis $\mathcal{B}$ for the CNS space. It is
worth noting that the different neuronal populations could be interdependent
(i.e., $\mathcal{B}$ does not represent an orthonormal basis), and,
the CNS is a nonlinear system, so that even small variations of the
activity (or the elicitation) of a single response component could
lead to a very different outcomes (for instance, a small variation
of the mean firing rate of a population, or of the quantity of a given
neurotransmitter, can even lead to an opposite behavioral outcome;
\citep{Goodman2006}).

\emph{Reactive mimicking. }In a growing body of literature it has
been reported that pharmacological conditioning (and implicit UCS
revaluation) determines reactive responses which mimic the original
active pharmacological effects. For instance, experimental results
reported in \citep{Ito2000} show that an increase in dopamine release
in the \emph{ventral striatum}, measured through microdialysis, are
observed not only when rats self administer cocaine (UCS), but also
when they are solely presented with a tone (CS) that has been previously
paired with cocaine administration. Furthermore, in \citep{Amanzio1999}
it is shown that pharmacological conditioning, like conditioning with
opioids, produces placebo analgesia mediated via opioid receptors,
and that, if conditioning is performed with nonopioid drugs, other
nonopioid mechanisms are involved, so that conditioning activates
the same neuronal populations as the active pharmacological treatment.
Similar results and conclusions have been obtained in mice experiments
\citep{Guo2010}. Moreover, brain imaging data, evidence that placebos
can mimic the effect of active drugs and activate the same brain areas;
this occurs for placebo-dopamine in Parkinson's disease \citep{DeLaFuente2001,DelaFuente2002},
for placebo-analgesics \citep{Amanzio1999,Eippert2009,Guo2010,Lui2010,Nolan2012,Petrovic2002}
or antidepressants, and for placebo-caffeine in healthy subjects \citep{Haour2005}.
It is worth noting that since the active pharmacological response
is highly differentiate, the reactive mimicking property leads to
the highly differentiation of the \emph{reactive system}. Such a system
can be thought as a distributed network within the CNS, which generates
reactions (or self-made responses) to relevant stimuli and substances. 

The above mentioned properties of pharmacological responding, that
are, integration, differentiation and reactive mimicking, can be described
from a quantitative perspective; this should lead to the derivation
of the laws governing the origin and the dynamics of the placebo response.
To this aim it is also necessary describing the dynamics of the UCS
revaluation, that is the implicit UCR evaluation over successive administration
trials.
\end{singlespace}

\section*{\noindent Error-Driven Learning}

\noindent From a growing body of literature emerges that learning
occurs through the computation of specific \emph{error-signals }(or
\emph{prediction errors}) (see \citep{Garrison2013,schultz2000b}
for a review). Generally speaking, the prediction error is defined
as the difference between the response expected from a given stimulation
and the response actually perceived by the elicited organism. This
definition relies on experimental observations acquired in functional
imaging studies \citep{Berns2001,Doherty2003,Garrison2013}, or directly
measured in dopaminergic circuits (e.g., in the \emph{ventral tegmental
area}, VTA) or in other fear-related circuits \citep{Bray2007,Delgado2008b,Li2014,McNally2011,Schultz2000,schultz2000b,Schultz2006,Steinberg2013,Waelti2001}.

\begin{singlespace}
Different mathematical models describing classical conditioning learning
(e.g., Rescorla-Wagner model \citep{Miller1995,Rescorla1972} or \emph{temporal
difference }(TD)\emph{ models} \citep{Doherty2003,Schultz1997,Sutton1988,Sutton1990}),
or describing learning in general, such as the probabilistic (Bayesian)
``perception'' and ``action'' learning models (i.e., the \emph{predictive
coding }(PC) \citep{Friston2003,Friston2008} and the \emph{active
inference} \citep{friston2009,friston2010}), assume that coding behavioral
responses involves the computation of an error-signal. More specifically,
the brain makes predictions in relation to a given stimulus and, on
the basis of the experienced outcome, the prediction is updated through
the prediction error. If the experienced outcome is greater (lower)
than the prediction, the computed error signal is positive (negative)
and corrects the new prediction; furthermore, if the experienced response
coincides with the expected outcome, the error signal is zero and
no prediction updatings take place. 
\end{singlespace}

\begin{singlespace}

\section*{\noindent Theoretical Concepts and Quantities}
\end{singlespace}

\emph{Differentiation.} Eq. (\ref{eq:base_model}) describes the elicited
response within the CNS accounting for all the involved neuronal populations
(or response components); however, we can consider one single component
to ease the reading. This choice, however, does not entail any loss
of generality, since our model can be applied to any component.

\begin{singlespace}
\emph{Integration}. As previously shown, the response to a drug administration
can be expressed by three main contributions (each of which might
be further decomposed in more detail), these are: 1) the active pharmacological
response, denoted $x$; 2) the reactive (self-induced) response due
to implicit UCS revaluation learning and classical conditioning, denoted
$i_{R}$; 3) the reactive response due to expectations and, generally
speaking, to higher cognitive information processing (such as beliefs,
social observations and verbal suggestions), denoted $\gamma$. Hence,
the integration property leads to the following expression for the
general elicited neuronal population:

\begin{equation}
y=x+i_{R}+\gamma.\label{eq:integration_property}
\end{equation}

It is important to note that the above described ``response decomposition''
is only a theoretical abstraction, since, because the integration
property, the response is unique and the brain cannot discriminate
between the different contributions. 

\emph{Reactive mimicking.} Since the reactive response ($i_{R}$)
associated with a given substance determined by implicit UCS revaluation
or by pharmacological conditioning mimics the active pharmacological
effect ($x$), the reactive response for a given neuronal population
can be expressed as:

\begin{equation}
i_{R}=\alpha\cdot x,\label{eq:ir-first-trial}
\end{equation}

where the term $\alpha$ takes on real values and represents the \emph{efficiency}
(or the \emph{strength}) of the reactive system for the given neuronal
component.

\emph{Implicit UCS revaluation.} If multiple successive open drug
administration trials are considered, provided that no conscious information
are available (such as verbal suggestions or beliefs) about the drug
effects, it can be assumed that in every trial the expected (or predicted)
response coincides with the last experienced outcome (wich, in turn,
coincides with the response experienced in the previous trial), or,
alternatively, that the expected response converges over successive
trials to the actual experienced outcome. Without any loss of the
generality, we can assume that the predicted outcome is equal to the
last experienced outcome. Provided that the active pharmacological
effect ($x$) is constant over successive drug administration trials,
and that the higher cognitive information processing are avoided (in
other words, $\gamma=0$, see Eq.(\ref{eq:integration_property})),
the reactive response due to implicit learning (i.e., due to the UCR,
or $y$ evaluation and revaluation over trials), $i_{R}$, is updated
through the error-signal computation. In the first trial the reactive
response is equal to zero, since no previous learning occured for
the given drug administration (UCS); hence, $y_{1}=x$. Since the
expected outcome was equal to zero for that UCS, the error signal
after the first administration trial is equal to $x$ and, such a
prediction error updates the reactive response, according to Eq. (\ref{eq:ir-first-trial}).
In the second open drug administration trial the response is given
by Eq. (\ref{eq:integration_property}) with $\gamma=0$, that is
$y_{2}=x+\alpha x$; moreover, a new error signal is computed as $y_{2}-y_{1}=\alpha x$,
and the reactive response updated at the end of the second trial is
given by $i_{R2}=\alpha x+\alpha^{2}x$. It easy to demonstrate (see
Supplemental Information) that the response elicited in the n-th drug
administration trials can be expressed as:

\begin{equation}
y_{n}=x+x\sum_{k=1}^{n-1}\left(\alpha^{k}\right).\label{eq:yn}
\end{equation}
 Eq. (\ref{eq:yn}) shows that the response to a given administration
drug diverges (to infinity) or converges to an asymptotic value, as
the number of trials increases, depending on the magnitude of $\alpha$.
The experimental observations (and the ordinary life experiences)
show that the pharmacological responding tends to reach an asymptotic
value over successive trials and does not diverge; hence the absolute
value of $\alpha$ has to be comprises between zero and the unity
(i.e., $0<\left|\alpha\right|<1$). We denote the above mathematical
condition as \emph{stability of the reactive system.} Under such a
condition it is easy to demonstrate that the asymptotic response can
be expressed as follows:

\begin{equation}
y_{\infty}=x+x\cdot\frac{\alpha}{1-\alpha},\label{eq:asymptotic_value}
\end{equation}
where the active pharmacological effect ($x$) and the reactive component
learned during successive trials (i.e., the quantity given by $i_{R\infty}=x\cdot\frac{\alpha}{1-\alpha}$),
are highlighted. At this stage, if in a successive trial an inerte
substance is administered (in other words, if a placebo is given,
so that the active pharmacological effect is brought to zero, $x=0$),
the placebo response coincides with the reactive response $i_{R\infty}$,
implicitly learned over trials. We argue that such a reactive response
represents the \emph{unconscious }or\emph{ implicit placebo response}.
If the placebo is administered during successive trials, it is easy
to demonstrate that the response tends asymptotically to zero, driven
by the error signal (see Supplemental Information). It is important
pointing out that the open administration (UCS) allows the individual
to perceive the UCS (i.e., the drug administration itself), and hence
to trigger the learned reactive response associated with that UCS;
furthermore, an error signal is computed whenever the UCS outcome
is revaluated. Conversely, if the administration was hidden, no UCS
can be perceived and no reactive response can be triggered. Moreover,
it is arguably that different features of the drug administration
(or substance intake), such as the flavor of a given pill (or even
conditioned stimulus), contribute to perception and differentiation
of the given substance intake (UCS). 

Finally, the \emph{integration, }the \emph{reactive mimicking }and
the \emph{stability of the reactive system }properties lead to the
following theorem:
\end{singlespace}
\begin{thm}
\begin{singlespace}
It is necessary that two distinct quantities are encoded within the
CNS for the stability of the reactive system, these are the reactive
response and the predicted (expected) outcome.

\emph{The demonstration is given on the Supplemental Information.
In practice, the theorem proves that both the reactive response ($i_{R}$)
and the expected (or predicted) outcome have to be computed and independently
updated in order to assure the stability of the reactive system. In
other words, the theorem proves that the reactive response cannot
be represented by the predicted (i.e., expected) response.}\end{singlespace}

\end{thm}
\begin{singlespace}

\section*{Outcome Simulation and Conscious Placebo/Nocebo Response}
\end{singlespace}

Since the above mentioned theorem proves that both the reactive response
and the expected outcome have to be coded within the CNS, the response
to drug administration (more specifically the reactive component of
it) should be a function of these quantities. In the previous Section
it is shown how, in the absence of cognitive information processing,
the reactive response by UCS revaluation learning through the computation
of the error signal. In this Section it is proposed how it is possible
modulating the reactive response through cognitive and conscious information
processing, such as verbal suggestions, beliefs, expectations, and
even by social observational learning \citep{Colloca2009}. More specifically,
we hypothesize that cognitively processed information, which permits
to infer (i.e., \emph{to simulate}) the outcome of a given substance
intake, lead to a process that we denoted \emph{outcome simulation}.
Such a process consists of high level cognitive processing of different
pieces of information related to the outcome, and lead to the computation
of the most probable outcome. Certainly such a computation depends
also on the previous experiences which are evoked by the provided
information. If the simulated outcome is different from the expected
outcome (i.e., from the previously experienced outcome) an error signal
is computed and the reactive response ($i_{R}$) is updated accordingly.
From a mathematical and modeling perspective, the outcome simulation
process corresponds to a \emph{virtual }substance administration trial,
in which the experienced outcome is the simulated (predicted or inferred
through cognitive processing) outcome, and the error signal is computed
as the difference between the simulated and the expected. It easy
to prove that the reactive response is updated by the error signal
through the following recursive expression \citep{Puviani2016}:

\begin{singlespace}
\begin{equation}
i_{R,n}=i_{R,n-1}+\alpha\cdot e_{n},\label{eq:updating_ir1}
\end{equation}
where $e_{n}$ represents the error signal computed in the n-th trial
(real or virtual), provided that $\alpha$ does not change over time;
otherwise, the more general expression for $i_{R}$ in the n-th trial
is as follows:

\begin{equation}
i_{R,n}=\alpha_{n}\sum_{k=1}^{n-1}e_{k}.\label{eq:updating_ir2}
\end{equation}

For these reasons, verbal suggestions or beliefs which let a given
individual inferring a positive (negative) outcome, lead to the computation
of a positive (negative) error signal which increase (decrease) the
reactive response associated with a given drug or treatment. We argue
that the simulation process takes place not only in drug administration
but even in more general situations, such as in stimulus-outcome learning
(or UCS evaluation); for instance animals could learn that a given
stimulus is dangerous observing others facing with such a stimulus
\citep{Olsson2007}. Moreover, cognitive information processing could
influence the reactive response associated with different stimuli
and substances; for instance, in \citep{Plassmann2008} is shown,
by functional MRI studies, how prices of wine bottles (which represent
here a piece of information related to the outcome) can affect the
experienced pleasantness of the wine intaking (UCS). Indeed, the experienced
pleasantness (which represents here the UCR) is due to the integration
of the active component, $x$, and the reactive (self-induced) response
$i_{R}$, which is due to prior learning and/or cognitive evaluation.
Concluding, cognitive information processing could lead to an outcome
simulation process which, in turn, modulates the reactive response
through the computation of an error signal, uderstood as the difference
between the simulated (inferred) outcome and the expected (previously
experienced) outcome.
\end{singlespace}

\begin{singlespace}

\section*{\noindent Operational Measures and Prediction of Implicit Placebo
Response}
\end{singlespace}

\noindent Given a specific target neuronal population, the active
response to a pharmacological administration ($x$) could vary across
individuals, and, more important, the reactive response component
(i.e., the implicit placebo or reactive response, $i_{R}$) could
be even more variable. Indeed, the parameter $\alpha$, which drives
the reactive dynamics, should be a function of different parameters,
such as the specific neural population involved (such as dopaminergic,
opioidergic, serotonergic, and so on) and the number, the types and
the sensitivity of the receptors within the given population. In turn,
these variables depend on different parameters, for instance these
are functions of the individual\textquoteright s genetic makeup. Indeed,
there is growing evidence that the individual\textquoteright s genetic
makeup (a stable trait) influences clinical outcomes and potentially
may allow for identification of placebo responders (see \citep{Hall2015}
for a review). More specifically, treatment outcomes in the placebo
arms of trials that have assessed genetic variation in the dopaminergic,
opioid, cannabinoid, and serotonergic pathways suggest that genetic
variation in the synthesis, signaling, and metabolism of these neurotransmitters
contributes to variation in the placebo response \citep{Hall2015}.
For these reasons, it could be useful to estimate (i.e., measure)
the quantities $x$ and $\alpha$ for a given pharmacological target
and for a specific genetic makeup. With such measures it should be
possible to optimize the combination \emph{drug-genetic makeup} and
considering the implicit placebo response as a stable and reliable
contribution of the pharmacological responding and, hence, as an indistinguishable
part of the ``true''\emph{ pharmacological treatment effect}. Operationally,
the measure of the active pharmacological response can be assessed
by hidden or covert drug administration. The response should be assessed
quantitatively adopting, for instance, functional brain imaging or
neurons activity recording, estimating the activity (understood as
the product of the mean number of firing neuron and the mean firing
rates) increase of the target neuronal population. Furthermore, the
parameter $\alpha$ can be assessed by open (overt) drug administration,
provided that the subject does not know which is the pharmacological
target \citep{Amanzio1999} (in other words the reactive response
due to implicit UCS evaluation is considered, and the expectations,
beliefs of clinical benefit or higher cognitive information processes
are avoided; i.e. $\gamma=0$, see Eq. (\ref{eq:integration_property})).
More specifically, assessing the increase of the given neuronal population
activity in successive drug administration trials permits to estimate
the increase of the overall response (which comprises the active and
the reactive response) which will converge to an asymptote, according
to Eq. (\ref{eq:asymptotic_value}). Moreover, the asymptotic convergence
to zero of the activity of the given neuronal population can be assessed
over successive placebo administration trials. Hence, the parameter
$\alpha$ could be reliably estimated from these data through simple
mathematical computations.

\begin{singlespace}
Moreover, since an important prediction of the model is the dynamics
of the error signals and of the reactive responses over successive
open (active and placebo) administration trials, these quantities
can be experimentally measured or estimated. In particular, experimental
measures obtained recording dopamine neurons (e.g., see \citep{Schultz1997,Schultz1998,Schultz2000,schultz2000b,Schultz2006,Waelti2001})
reveal that dopamine neurons in specific brain regions (such as the
ventral tegmental area, VTA, or the substantia nigra) code prediction
and prediction error signals related to rewards (in particular related
to rewarding substances, drugs or food delivery). Despite these experiments
were designed to study dopamine neurons activity in classical conditioning
and in the unexpected (and predicted) delivery of rewarding susbstances,
these suggest the possibility to monitor the activity of the VTA dopamine
neurons (for instance in rats) during the administration of drugs
which determine a rewarding response, such as cocaine or morphine.
In particular, it is expected that in hidden drug administration trials
no phasic activity of VTA dopamine neurons has to be observed; moreover,
the same result should be observed during the first open drug administration
trial. Furthemore, starting from the second administration trial,
the coding of the reactive response associated with the drug administration
(UCS) has to be observed, and such a response should increase asymptotically
over successive trials. Finally, the VTA dopamine neurons activity
should tend asymptotically to zero over successive placebo administration
trials. From the above mentioned experimental measures it is possible
to estimate the $\alpha$ parameter for the dopaminergic neuronal
population related to a specific rewarding drug (indeed, different
rewarding signals and neuronal populations exist, related to liquid
or solid rewarding substances, drugs, and other rewarding stimuli;
\citep{Schultz2000}). It is worth pointing out that the VTA neurons
activity have to be monitored for all the duration of the drug effect
within each trial, since it is arguably that the reactive response
mimicks also the temporal trend of the active drug effect. In fact,
in the above mentioned measures performed on the VTA neurons, time
represents a fundamental variable which is coded by neurons (since,
for instance, the time of the reward occurence plays and important
role). In other words, it should be possible that, if the active drug
effect starts after a time delay from the administration, and it vanishes
with a specific temporal trend, the reactive response could mimick
the same temporal trend.
\end{singlespace}

\section*{Obtaining Increasing and Resistant-to-Extinction Reactive Responses}

\noindent Since the mammalian brain is able to induce reactive responses
similar to those obtained by pharmacological active effects, exploiting
the implicit reactive system dynamics for obtaining a desired outcome
should be feasible, in principle. More specifically, our system-level
model suggests that it could be possible obtaining resistant-to-extinction
reactive responses manipulating the error signal computation. Indeed,
even avoiding and neglecting cognitive processing (e.g., verbal suggestions)
during drug administration or substance intake, as previously shown,
the error signal is implicitly computed, and it drives (or updates)
the reactive response $i_{R}$ associated with the representation
of the given drug (i.e. the UCS; see also \citep{Puviani2016} for
a more comprehensive analysis). Provided that, for a given genetic
makeup and for a specific target CNS component, the reactive system
efficiency $\alpha$ is greater than zero, after some active drug
administration trials the reactive response reaches the asymptote
$i_{R\infty}=x\cdot\frac{\alpha}{1-\alpha}$ (see Eq. (\ref{eq:asymptotic_value}))
through implicit UCS revaluation learning, and, hence, through the
successive updatings of the reactive response by error signal computations
(see Eqs. (\ref{eq:updating_ir1}-\ref{eq:updating_ir2})). Furthermore,
if a placebo administration trial follows, then the implicit placebo
response, which is equal to the asymptotic value $i_{R\infty}$, is
experienced, moreover, the experienced outcome is updated and an error
signal is computed, leading to a decrease of the reactive (placebo)
response, which vanishes over successive placebo trials. However,
if the error signal computation is blocked or ``degraded'' during
placebo (or lower active dose) administration, the experienced outcome
is updated; conversely the reactive response (and hence the implicit
placebo response) does not decrease (or decrease by little), since
no error signal (or a degraded error signal) updated it. Furthermore,
a resistant-to-extinction reactive response can be obtained, since
no error signals can be computed over successive trials, because the
expected outcome concides with the reactive response, which, in turn,
coincides with the experienced outcome; in other words, a reactive
response which does not tend asymptotically to zero over successive
placebo trials is obtained. We denote such a response $i_{R0}$, since
it will be elicitated by the reactive system at every substance intake
trial (as it were originally generated by a fictitious trial \emph{zero}),
regardless of whether an active pharmacological effect exists. Indeed,
it is easy to demonstrate that, if a resistant-to-extinction reactive
response $i_{R0}$ has been associated to a given substance, the expression
for the response to substance intake driven through implicit learning
in the generic n-th trial is\footnote{In \citep{Puviani2016} it is shown that different resistant-to-extinction
emotional reactive responses are natively coded in the mammalian brain
from birth, and these are associated with phylogenetic (i.e., prepared
biological and evolutionary fear relevant) stimuli representations
\citep{Esteves1994,Ohman1993}. }:

\begin{singlespace}
\begin{equation}
y_{n}=i_{R0}+x+\alpha\left(y_{n-1}-i_{R0}\right).
\end{equation}

The resistant-to-extinction property of a reactive response is the
key element for increasing unlimitedly an implicit response; indeed,
repeating the cycle of trials (i.e., the above described protocol)
adopted for obtaining $i_{R0}$, will lead to the increase of the
reactive response, through an accumulation principle (see Supplemental
Information for the computational details). Hence, through multiple
cycles of active pharmacological drug administration followed by inerte
(or a lower active dose) substance administration, together with a
pharmacological treatment able to degrade the error signal computation
(more specifically able to lower the precisions associated with error
signals, see \citep{friston2009,Friston2008}), it is possible to
``force'' the CNS to react to such a drug with an increasing self-induced
response which mimics the active pharmacological treatment. After
some cycles it is expected that the reactive response $i_{R}$ will
be greater than the original active pharmacological response, and
the pharmacological effect can be increased indefinitely.
\end{singlespace}

In practice, it is expected that the error signal cannot be completely
eliminated in the placebo administration trials, so that multiple
placebo administration trials with an ``error signal degradation
treatment'' will be required, and the resulting $i_{R0}$ will be
smaller than $i_{R\infty}$. Nevertheless, the proposed implicit learning
model describes the reactive system dynamics at a system (macro) level
and, even if it predicts the possibility of obtaining a resistant-to-extinction
(i.e., stable) and increasing reactive response, it does not give
any indications of how obtaining it, nor which pharmacological targets
have to be considered for the degradation of the error signal. For
these reasons, a specific model, which can describe the error signal
computation at a deeper level (i.e., at the neurons level) is required.
We argue such a model could be represented by the so called \emph{predictive
coding} model (and its version applied to action learning; \emph{active
inference}) \citep{Friston2003,Friston2008,friston2009,friston2010}.

\begin{singlespace}

\section*{\noindent Predictive Coding, Active Inference and Error Signal Computation}
\end{singlespace}

\noindent In predictive coding \citep{Friston2003,Friston2008} is
formalized the notion of the Bayesian brain, in which neural representations
in the higher levels of cortical hierarchies generate predictions
of representations in lower levels. These \emph{top-down }predictions
are compared with representations at the lower level to compute a
\emph{prediction error}. The resulting error-signal is passed back
up the hierarchy to update higher representations; this recursive
exchange of signals lead to the minimization (ideally the suppression)
of the prediction error at each and every level to provide a hierarchical
explanation for sensory inputs that enter at the lowest (sensory)
level. In the Bayesian jargon neuronal activity encodes beliefs or
probability distributions over states in the world that causes sensations
\citep{Friston2008}. In predictive coding the notion of \emph{precision}
(or confidence, which represents the inverse of the variance) of the
error signals is also formalised, and the mechanism through which
the brain has to estimate and encode the precision associated with
the prediction errors is explained. The prediction errors are then
weighted with their precision before being assimilated at a high hierarchical
level. More specifically, active inference accounts posit that the
brain deals with \emph{noisy }prediction errors by decreasing the
gain on cortical pyramidal neurons that function like precision units
to regulate outputs from signal error computations, thereby reducing
the influence of these outputs \citep{Barrelt2015,Friston2003,Pezzulo2015}.
It is important pointing out that the resulting error signal depends
on the relative precision of prediction errors at each level of the
neural hierarchy. Hence, a neuropharmacological alteration of the
precision on different levels in the cortical hierarchy could lead
to the modulation of the error signal computation. Furthermore, it
is supposed \citep{adams2013,Feldman2010} that such an alteration
may naturally occur in some pathologies, such as in Parkinson's desease,
where the depletion of dopamine (which is supposed to encode the precision
of prediction errors by altering their synaptic gain, as also other
neurotransmitters do \citep{Feldman2010}) at different levels, would
alter the balance of precision at higher (sensorimotor) relative to
lower (primary sensory) levels in the cortical hierarchy; moreover,
the symptoms vary according to the site (hierarchical level) of changes
in precision \citep{adams2013}.

\vspace{1cm}

\section*{\noindent Neuronal Populations within the Reactive System}

\noindent Which are the neuronal populations belonging to the reactive
system? In other words, for which CNS response components does the
reactive mimicking property hold? If a given neuronal population belongs
to the reactive system, then it is possible to obtain a reactive response
(i.e., an implicit placebo response) for this neuronal population.
Experimental results from the literature evidence that at least the
following CNS components could generate a reactive response: 1) emotional
system responses \citep{Puviani2016} (which include, for instance,
the dopaminergic mesolimbic and mesocortical system, \citep{Scott2007,Colloca2014a},
the endocannabinoid and opioid system in placebo analgesia, \citep{Eippert2009,Nolan2012,Pascalis2002,Petrovic2002,Wager2007,Watson2009,Zubieta2005},
the serotoninergic system, the target neuronal systems of depression,
anxiety and addiction; see \citep{Benedetti2008}); 2) the dopaminergic
motor system \citep{DeLaFuente2001,DelaFuente2002}; 3) the \emph{humoral
immune response system }(in particular the components of the CNS such
as the hypothalamic-pituitary-adrenal axis, HPA, or the sympathetic
nervous system, SNS; \citep{Benedetti2008,cacioppo2007,Goebel2002,Vits2011});
4) the endocrine system; (see \citep{Benedetti2008,Enck2008} for
a review). We speculate that the main components of the reactive system
are the emotional/motivational and the humoral immune systems, since,
from an evolutionary perspective, the organism has to be able to learn
and to automatically react whenever ``important'' stimuli are perceived,
and, implicitly recognize ``relevant'' substances, such as food,
in order to approach or avoid them or to quickly and automatically
react (for instance through an automatic trigger of a first immune
response). More specifically, a stimulus is ``important or relevant''
for an organism if it previously elicited a CNS response belonging
to the reactive system, such as a painful stimuli (which are related
to emotional components), a rewarding substance (such as food), or
primary immune responses.

\begin{singlespace}

\part*{\noindent Discussion}
\end{singlespace}

\noindent We propose a model describing response to substance intake
based on implicit UCS revauation learning (which describes the automatic
evaluation and revaluation of an unconditioned stimulus, that represents
the given substance intake). Such a model formalizes the key properties
of drug administration response based on well established empirical
evidence. The theoretical conceptualization leads to a theorem which
proves the necessity of the encoding of two fundamental quantities:
the reactive response and the expected (or predicted) outcome. We
have derived a model that meets this requirement and is able to predict
the response to drug administration over successive trials. The placebo
response is mathematically formalized as a reactive response, which
represents an implicit and inevitable component of the overall response
to substance intake; moreover its effect becomes naturally evident
when an organism responds to a substance administration whose active
(pharmacological) effect has been eliminated.

\begin{singlespace}
Modeling the dynamics of the reactive system, from which the placebo
response originates, permits also the definition of quantities (e.g.,
the reactive efficiency, $\alpha$) and of operational measures which
could lead to the maximization and the stabilization of the reactive
response ($i_{R}$) for a given genetic makeup, even when the pharmacological
active effect ($x$) is kept constant. Furthermore, the model shows
that it should be possible to create a resistant-to-extinction reactive
response blocking (or degrading) the error signal computation in specific
trials. More specifically, after some trials in which the active drug
is administered in order to increase the implicit reactive response,
successive trials have to follow, in which placebo administration
is given together with a pharmacological treatment able to degrade
the precisions of the error signal (and hence the overall error signal
computation). The targets of such a pharmacological treatment have
to be specific neurotransmitters on precise level of the neural hierarchy,
depending on the target component of the original active drug administration;
moreover, predictive coding and active inference model provides the
theoretical and computational basis for the assessment of such targets. 

In this manuscript, the theoretical feasibility study of a strategy
for the unlimited increase of a given drug effectiveness has been
developed. Further computational and experimental studies are needed
in order to explore the mentioned strategy. 

We also argue that the generality of our model permits the modeling
of a broad range of phenomena and pathologies which involve the interaction
between an organism and a substance intake, such as, drug addictions
and eating disorders, and for the development of novel strategies
for decreasing pathological resistant-to-extinction reactive responses,
such as in phobias or post-traumatic stress disorders. 
\end{singlespace}

\part*{Supplemental Information}

\begin{singlespace}
In the following sections the equations describing the reactive system
dynamics (and, hence, the implicit placebo response dynamics) are
derived. It is assumed that no cognitive information processes occur
(in other words, verbal suggestions, beliefs and expectations on clinical
benefit are avoided); however, drug administration, or substance intake,
has to be ``overt'', so that the considered individual is aware
that the same substance is administered over successive trials (but
he/she does not know which is the pharmacological target of such a
substance). Furthermore, to ease the computations, it is assumed that
the expected (predicted) response, in a given k-th trial, coincides
with the last experienced outcome (i.e., $y_{expected,k}=y_{k-1}$).
\end{singlespace}

\begin{singlespace}

\section*{Implicit Response Acquisition to a Substance Intake}
\end{singlespace}

\begin{singlespace}
In this section the responses elicited during succesive trials in
which an active drug is administered are derived.

In Table 1 the variables of interest are mathematically described
through succesive trials. A single CNS component (i.e., a specific
target neural population) is considered for the computations, provided
that such a component belongs also to the \emph{reactive system}.

\includegraphics{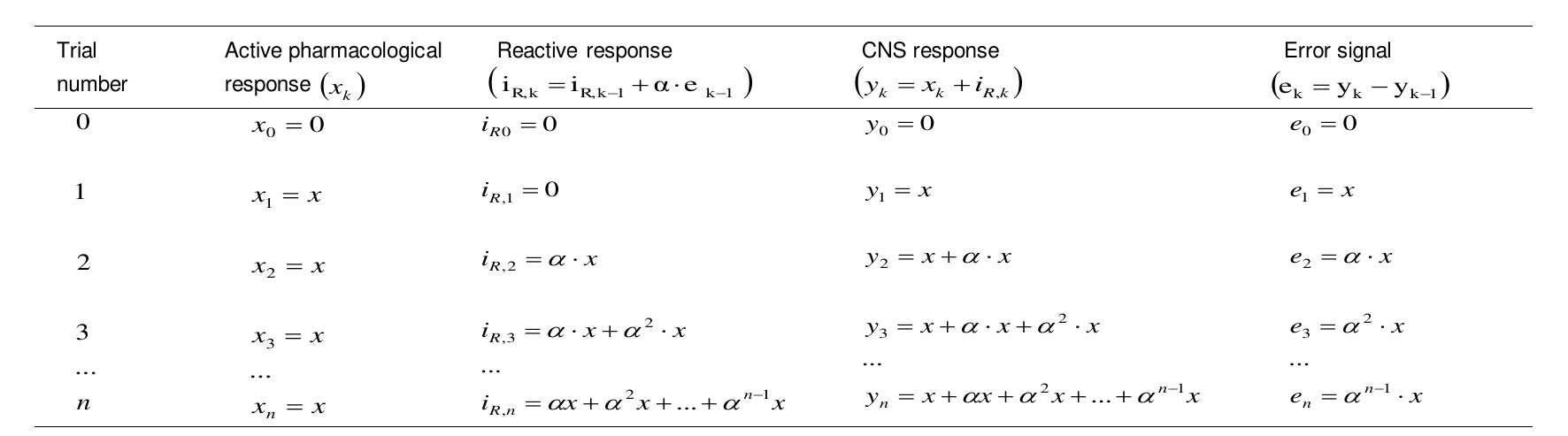}
\end{singlespace}

\noindent \textbf{\emph{\footnotesize{}Table 1. }}\textbf{\footnotesize{}In
this Table the computations leading to the automatic central nervous
system (CNS) response to a drug intake over successive trials are
derived. It is assumed that the active pharmacological effect ($x$)
is constant over successive trials.}{\footnotesize \par}

\begin{singlespace}
\vspace{1cm}

The response to drug intake in the n-th trial can be written as:

\begin{equation}
y_{n}=x+x\sum_{k=1}^{n-1}\left(\alpha^{k}\right)\label{eq:yn_acq_sum}
\end{equation}

Considering the reactive system stability property, it follows that
the term $\alpha$ has to be less than one in magnitude (i.e. $0<|\alpha|<1$).
Hence the summation term in eq. (\ref{eq:yn_acq_sum}) converges to
the value $\alpha/(1-\alpha)$, for the property of the geometric
series, when the number of trials tends to infinity. Thus, the asymptotic
response due to the active drug administration can be written as:

\begin{equation}
y_{\infty}=\frac{x}{1-\alpha}=x+\frac{\alpha\cdot x}{1-\alpha}.\label{eq:y_infinitive}
\end{equation}

The eq. (\ref{eq:yn_acq_sum}) can also be expressed as:

\begin{equation}
y_{n}=x+\alpha\cdot y_{n-1}\label{eq:yn_recursive}
\end{equation}

which shows the recursive nature of the response acquisition process
over successive trials.
\end{singlespace}

\begin{singlespace}

\section*{Reactive or Implicit Placebo Response Dynamics}
\end{singlespace}

\begin{singlespace}
In this section the response elicited during succesive trials where
a placebo is administered (i.e. the active pharmacological component
is absent) is derived.

In Table 2 the variables of interest are mathematically described
over succesive trials. A single generic component is considered in
the computations, moreover, it is assumed that the asymptotic response
has been reached over the previous active drug administration trials
(see Eq.(\ref{eq:asymptotic_value})).

\includegraphics{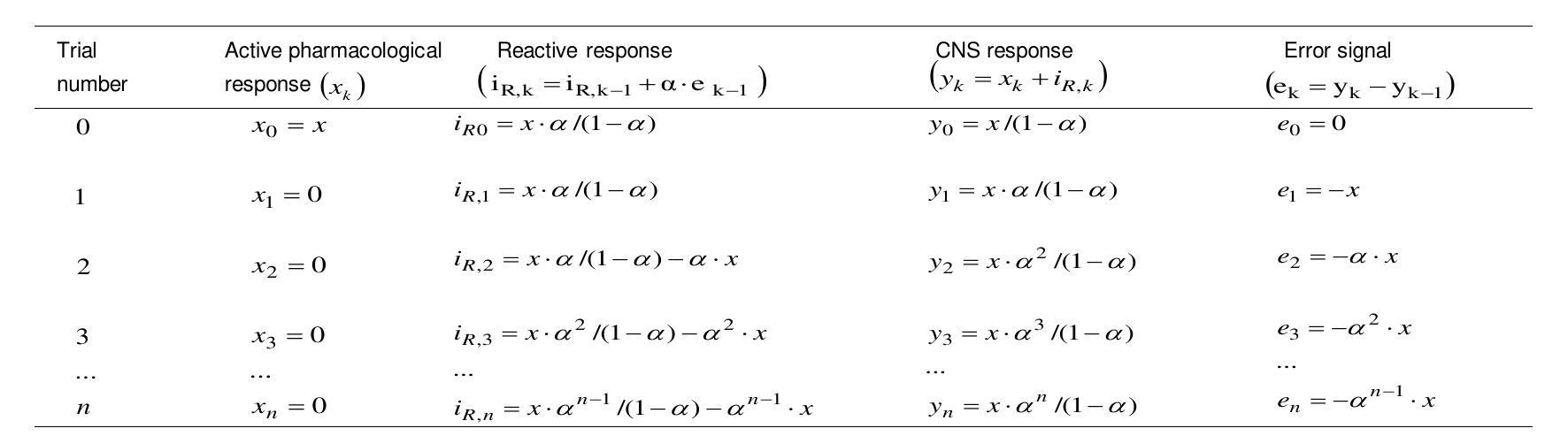}
\end{singlespace}

\begin{singlespace}
\noindent \textbf{\emph{\footnotesize{}Table 2. }}\textbf{\footnotesize{}-
In this Table the computations leading to the reactive (implicit placebo)
response extinction over successive trials are derived. It is assumed
that, previously to placebo admnistration trials, the pharmacological
effect was equal to $x$.}{\footnotesize \par}
\end{singlespace}

\begin{singlespace}
\vspace{1cm}

The response to placebo administration in the n-th trial can be written
as:

\begin{equation}
\begin{cases}
y_{n}=\alpha\cdot y_{n-1}\\
y_{0}=Y_{0}
\end{cases},\label{eq:yn_ext}
\end{equation}

where, provided that the asymptote had been reached during the previous
active pharmacological trials, $Y_{0}=x\cdot\alpha/(1-\alpha)$.

\textbf{Theorem}\emph{: It is necessary that two distinct quantities
are encoded within the CNS for the stability of the reactive system,
these are the reactive response and the predicted }(\emph{expected})\emph{
outcome.}
\end{singlespace}
\begin{proof}
\begin{singlespace}
\textbf{\emph{Proof by contradiction (reductio ad absurdum)\label{Proof-by-contradiction}}}\end{singlespace}

\end{proof}
\begin{singlespace}
\medskip{}

\emph{Hypothesis} 1: The reactive response associated with a given
substance intake coincides with the expected (predicted) outcome,
and the expected or predicted outcome converges to the experienced
outcome.

\emph{Hypotheses }2: The properties integration, reactive mimiking
and reactive system stability hold.

\emph{Hypotheses }3: Cognitive information processing are avoided
(in other words, no beliefs, expectations for clinical benefits, verbal
suggestions and so on, are involved).

\medskip{}

Hypothesis 1 asserts that a unique reactive signal predicting the
drug administration outcome exists, and that this signal coincides
with the reactive response elicited when the drug is administered;
furthermore, the predicting signal converges (by learning) to the
actual experienced elicitation. The last assumption has been formulated
to include a more general scenario than that considered in our initial
assumptions, in which the expected outcome coincides with the last
experienced outcome. From a mathematical viewpoint, the expected outcome
can be computed using any supervised learning method (or, alternatively,
TD methods \citep{Sutton1988}) in which the predicted outcome is
evaluated on the basis of the past $m$ predictions (i.e., of the
predicted outcomes in the last $m$ trials) and of the actual outcome,
minimizing the error between the prediction and the experienced outcome.
Otherwise it can be assumed that the predicted outcome coincides with
the last experienced outcome.
\end{singlespace}
\begin{enumerate}
\begin{singlespace}
\item Let the UCS be the drug administration.
\item Drug administration to a given subject takes place on successive trials,
so that it exerts an active elicitation (i.e., it elicits the active
response $x$). During the first trial the response is exclusively
due to the active pharmacological component, that is $y_{1}=x$. After
the first trial (for instance, during the drug administration in the
second trial), the predicted (reactive) response, called $y_{predicted,1}$,
is computed. 
\item In the second trial, after the drug administration, the predicted
outcome ($y_{predicted,1}$) adds up to the successive UCS active
elicitation, so that the outcome can be expressed as $y_{2}=y_{expected,1}+x$.
Furthermore, since $y_{predicted,1}$ does not coincide with the actual
experienced outcome, the new prediction $y_{predicted,2}$ is computed
after the second trial; it can be easily proved that $y_{predicted,2}>y_{predicted,1}$
(since the experienced outcome has been strengthened and the error
signal has to be minimized).
\item In the third trial the experienced outcome can be written as $y_{3}=y_{predicted,2}+x$;
since $y_{3}>y_{2}\geq y_{predicted,2}$ a new value for the predicted
response is computed, called $y_{predicted,3}$, such that $y_{predicted,3}>y_{predicted,2}$.
\item In the $n$-th trial the outcome can be expressed as $y_{n}=y_{predicted,n-1}+x$;
it is easy to prove that $y_{n}>y_{n-1}\geq y_{predicted,n-1}$. Moreover,
if the number of trials tends to infinity, the outcome grows indefinitely
(i.e., $\underset{n\rightarrow\infty}{\lim}y_{n}=\infty$).
\item The last statement is absurd, as it contradicts Hypothesis 2, in particular
it contradicts the stability system principle. \end{singlespace}

\end{enumerate}
\begin{singlespace}

\section*{Increasing a Reactive Response through Resistant-To-Extinction Increments}
\end{singlespace}

Provided that a resistant-to-extinction reactive response, denoted
$i_{R0},$ has been previously obtained (see Section ``Obtaining
Increasing and Resistant-to-Extinction Reactive Response''), it is
easy to show that the drug intake response is expressed as:

\begin{singlespace}
\begin{equation}
y_{n}=i_{R0}+x+\alpha\left(y_{n-1}-i_{R0}\right),\label{eq:after_1_cycle}
\end{equation}

where the term $i{}_{R0}$ represents the inextinguishable reactive
response obtained during the first protocol cycle. The generic response
at the n-th trial can also be expressed as:

\begin{equation}
y_{n}=i_{R0}+x+x\sum_{k=1}^{n-1}\alpha^{k}.\label{eq:second_form}
\end{equation}

Eqs. (\ref{eq:after_1_cycle}-\ref{eq:second_form}) can be obtained
from the computations shown in Tab. 3.

\includegraphics{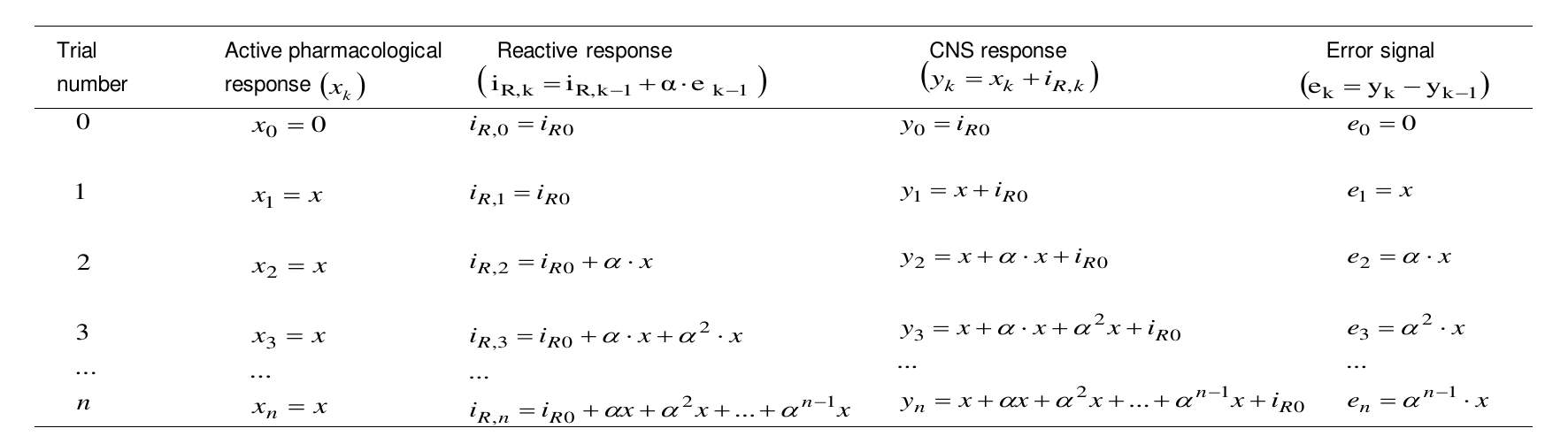}
\end{singlespace}

\begin{singlespace}
\noindent \textbf{\emph{\footnotesize{}Table 3. }}\textbf{\footnotesize{}-
In this Table the computations leading to the automatic central nervous
system (CNS) response to a drug intake over successive trials are
derived. It is assumed that the active pharmacological effect is constant
and equal to $x$, furthermore, a resistant-to-extinction reactive
response ($i_{R0}$) associated to the given substance intake has
been previously obtained, through the strategy described in the previous
sections. }{\footnotesize \par}
\end{singlespace}

\begin{singlespace}
\newpage{}

\textbf{\large{}References}{\large \par}

\bibliographystyle{vancouver/vancouver}
\bibliography{references_29-11sidita_APA}
\end{singlespace}

\end{document}